\documentclass[journal]{IEEEtran}
\usepackage{cite}
\ifCLASSINFOpdf
   \usepackage[pdftex]{graphicx}
\else
   \usepackage[dvips]{graphicx}
\fi
\usepackage{amsmath}

\usepackage{amsfonts}
\usepackage{amssymb}
\usepackage{tabularx}
\usepackage{caption}
\usepackage{subcaption}
\usepackage{algorithmic}
\usepackage{physics}
\usepackage{array}

\newcounter{MYtempeqncnt}

\begin{document}
\title{Adaptive Constellation Multiple Access for Beyond 5G Wireless Systems}

\author{Indu L.~Shakya, and Falah H.~Ali,~\IEEEmembership{Senior Member,~IEEE}
        % <-this % stops a space
\thanks{Manuscript received xxxxxxxxxxx y, 2023; revised xxxxxx y, 2023. The editor coordinating the review of this letter was xxxxxxxxxxxxxxxxx. (Corresponding author:
Indu L. Shakya.)

Dr. Indu L. Shakya was with the University of Sussex, Brighton BN1 9QT, U.K. He is now with e2E Group, Welwyn Garden City, AL7 1LT, U.K. (e-mail: ishakya@gmail.com). Falah H. Ali is with the School of Engineering and Informatics, University of Sussex, Brighton, BN1 9RH, UK. (F.H.Ali@sussex.ac.uk).}}
% and  the end of the author line. i.e., if you had this:

\markboth{IEEE Template for Letters,~Vol.~xx, No.~xx, yy~202x}%
{Shell \MakeLowercase{\textit{et al.}}: Bare Demo of IEEEtran.cls for IEEE Journals}



\maketitle

\begin{abstract}

We propose a novel non-orthogonal multiple access (NOMA) scheme referred as adaptive constellation multiple access (ACMA) which addresses key limitations of existing NOMA schemes for beyond 5G wireless systems. %Unlike the latter, which show degraded performances when power, modulation of grouped co-channel users are not within expected ranges, ACMA operates under much wider ranges with high performance, easing network deployment tasks. 
Unlike the latter, that are often constrained in choices of allocation of power, modulations and phases to allow enough separation of clusters from users' combined signals; ACMA is power, modulation and phase agnostic\textemdash forming unified constellations instead\textemdash where distances of all possible neighbouring points are optimized. It includes an algorithm at basestation (BS) calculating phase offsets for users' signals such that, when combined, it gives best minimum Euclidean distance of points from all possibilities. The BS adaptively changes the phase offsets whenever system parameters change. We also propose an enhanced receiver using a modified maximum likelihood (MML) method that dynamically exploits information from the BS to blindly estimate correct phase offsets and exploit them to enhance data rate and error performances. Superiority of this scheme\textemdash which may also be referred to as AC-NOMA\textemdash is verified through extensive analyses and simulations. 
%As the adaptation is not for instantaneous channel fading changes, signalling overheads are also significantly low.
\end{abstract}


% \fi
\IEEEpeerreviewmaketitle

\section{Introduction}
%\IEEEPARstart{T}{his}
% demo file is intended to serve as a ``starter file''
%for IEEE journal papers produced under \LaTeX\ using
%IEEEtran.cls version 1.8b and later.
\IEEEPARstart{N}{OMA} is a promising method to share bandwidth among multiple users without subdivision in time or frequency resources \cite{IEEElee:nomaser}-\cite{IEEEkara:powqam}. Power-domain (PD)-NOMA is a well-known variant that relies on channel gain disparity of users and successive interference cancellation (SIC) method to separate co-channel users' data to give higher sum rates than orthogonal multiple access (OMA) like OFDMA \cite{IEEElee:nomaser}, \cite{IEEEmakki:noma}. 
%Code-domain NOMA scheme \cite{IEEEniko:scma} such as space code multiple access also exists, that uses short spreading codes and highly complex iterative receiver design to allow more users than available time or frequency resources. 
%RSMA \cite{EURAmao:rsma} is a generalisation of NOMA which relies on the knowledge of transmit channel state information (CSIT) to allow the BS to adjust powers of users while still using the SIC detection method. 
The SIC method used in PD-NOMA can be sensitive to co-channel users' signal power ratios \cite{IEEEzhang:reliablenoma}. When the ratios are not optimized, it is unable to decode the interference and ends up with error floor\cite{IEEElee:nomaser}. Also general assumptions in the NOMA literatures like high channel gain disparity of users may not be applicable in many practical wireless environments. 
As such, conventional NOMA is not yet proven to provide sustained communication to be included as work items of 3GPP standards \cite{IEEEmakki:noma}. 
%Although the issues of SIC can be addressed to some extent by adopting a joint detection (JD) method, receiver alone does not solve the key problems of NOMA \cite{IEEEguan:phjd}. Therefore, it is highly desirable to %find a novel approach to transmit multi-user signals in ways to ensure separability of co-channel data regardless of users being allocated similar or disparate powers. 

The use of transmit signal phase rotation for NOMA to help improve detection using SIC and other methods is adopted in \cite{IEEEguan:phsic}-\cite{IEEElin:rotmd}. The scheme in \cite{IEEEqiu:aljsic} proposes to use phase rotation as a mean to increase time diversity gain assuming block fading of underlying channels while using single-user receivers. It is known that the SIC outperforms single-user receivers when users' channel gains are disparate. The work \cite{IEEEguan:phsic}, \cite{IEEEwang:rotsic} \cite{IEEElin:rotmd} use phase rotation optimization while using SIC. However, SIC method treats other users' signals as noise. So, their gains are limited when users have similar channel gains. The phase rotation-based schemes addressing limitations of SIC with joint detection (JD)\textemdash also referred to here as JD-NOMA\textemdash are proposed in \cite{IEEEguan:phjd}, \cite{IEEEchang:phsic}, \cite{IEEEng:pmpe}. But they are limited to two users with [QPSK, QPSK] or [QPSK, 16-QAM] setups and limited power allocation ratios of users due to strict rules to cluster constellation points within the boundary of I/Q axes. If higher rotations are imposed, the receivers cannot easily separate data as their detection boundaries become very complicated. 
%As their phase rotation mechanism avoid crossing of constellation points to other quadrants, limiting rotation angles to small values, they unlikely achieve the best minimum distance. 

We solve issues of NOMA schemes using a power, modulation and phase agnostic ACMA; where both transmission and detection systems work together to minimize probability of error for users. We devise a novel algorithm at the BS that forms unified composite signals designed to maximize the distance of each neighbour points rather than clusters. Here users' signal phases are blind adaptively rotated up to the full potentials utilizing already available knowledge of powers and modulations. Unlike JD-NOMA, it is not limited to small set of power and modulation choices only, making it suitable for diverse channel gain conditions. We also propose an effective MML based receiver to capitalize on the synergy provided by the BS and parameters it shared via control channels to blindly find and apply correct phase offsets to give best possible data rate and error performances. We prove superiority of ACMA through extensive analyses and simulation results.  

% To the best of our knowledge, no such new system has been introduced before that is simple in design and yet very effective. The ACMA approach can also be easily incorporated to extend the capacity of existing NOMA schemes. This can deliver the capacity promise of these potential 6G multiple access techniques in practice and deployment for future systems. 

%The gain of PD-NOMA was investigated in multiple studies \cite{IEEEwei:constr}-\cite{IEEEwalid:groupnoma}, \cite{3GPP:must}, \cite{3GPP:rel16noma}. A survey paper on NOMA \cite{IEEEmakki:noma} state that the %gains of NOMA over OMA across many 3GPP studies were found to be marginal; therefore it is still not included as work-item in the the 3GPP and left possible inclusion in beyond 5G. 
%Here transmitting side aids receivers by sharing the knowledge of users' modulation symbol alphabets and power information so that they do not treat each others' data as interference or noise \cite{Isha:colla}.
\vspace{-3mm}
\section{Signal and System Models}
We consider a downlink system with $K$ users served simultaenously by a base-station. The constituent modulated symbol of each user $x_{k}$ is taken from a square M-QAM constellation with symmetric in-phase (I) $i_{k}$ and quadrature (Q) $q_{k}$ components as follows: $x_{k} = i_{k}+\sqrt{-1}q_{k}$. Note that the ACMA approach is about intelligent use of constellation information and as such, it is generic enough to be used for both uplink and downlink using diverse modulation methods like M-QAM, M-PSK. Each user's symbols have alphabet of size $M_{k}$. The symbols are then multiplied by modulation specific scaling factor to ensure unity expected energy $\mathbb{E}\{\abs{{x}}^{2}\}=1$. The BS then applies user specific power allocation coefficients $\alpha_{k}$ to suit power budget for various conditions. This is also done to ensure that the ACMA composite signals maintain the same average and peak power constraints as in single-user case for fair reference i.e. $P_{T}=\sum_{k=1}^{K}P_{k};P_{k}=\alpha_{k}P_{T}$. The composite constellation of $K$ users $s$, transmitted from the BS can be written as:  
\begin{equation}
\label{eqn_s}
s = \sum \limits_{k=1}^{K}\sqrt{\alpha_{k}P_{T}}x_{k} \exp(j\theta_{k}+\delta_{k}),
\end{equation}
where $\theta^{k}=\angle(x_{k})$ is an initial phase angle of a user's original M-QAM signals and $\delta_{k}$ is a user specific phase offset coefficient. %In ACMA $\delta_{k}$ is calculated at the BS such that a composite signal  $s$ is formed to maximise the mimimium Euclidean distance of all constellation points. 
We use a baseband model of users'  radio propagation channels to elucidate key differences of interference mitigation mechanisms of different NOMA transceiver algorithms. We assume all schemes apply the same approach to compensate for the large and small scale path losses to aid the detection of user' data by adjusting the parameters $\alpha_{k},k=1,2,..,K$ where $0\leq \alpha_{k}P_{T}\leq P_{T}, \forall k$. The signal at the receiver of user $k$ can be written as:$r_{k} = h_{k}s + z_{k};  k=1,2,..,K$, where $h_{k}$ represents baseband model of complex fading channel between the BS and user $k$. We also assume independent Rayleigh distribution for fading channels with $\mathcal{CN}(0,1)$. Finally, $z_{k}$ represents the additive white Gaussian noise (AWGN) component with $\mathcal{CN}(0,N_{0})$.  
\vspace{-1mm}

\section{Proposed Transceiver Designs}

%The data estimation process at the receiver involves signal processing to arrive as close as possible to the original data $\hat{x}_{k} \rightarrow x_{k}, \forall k$. 
%In the following, we first highlight the features of NOMA SIC method first to compare and contrast with our %collaborative receiver.

\begin{figure}[!t]
\centering
\includegraphics[width=3.5in, height=1.6in]{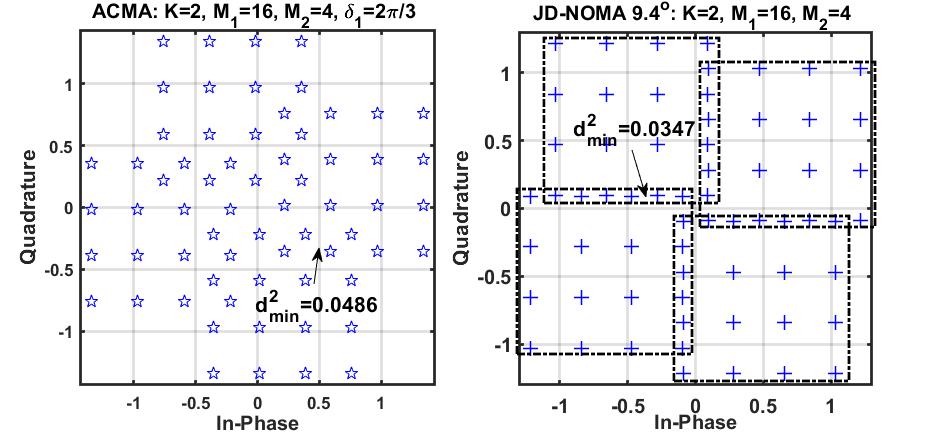}
\caption{Constellation of a) proposed ACMA and b) JD-NOMA \cite{IEEEguan:phjd}, \cite{IEEEchang:phsic} clusters for a two-user system using 16-QAM and 4-QAM and power allocation factors of $[\alpha_{1},\alpha_{2}]=[0.35,0.65]$. Phase offsets of $\delta_{1}=2\pi/3$ by the former and $9.4^{\circ}$ by the latter are generated, leading to different $d^{2}_{min}$.}
% ACMA gives higher $d^{2}_{min}$ as its phase offset values are not limited to certain constraints.}
\label{fig_const}
\vspace{-4 mm}
\end{figure}
 
\subsection{Conventional PD-NOMA Transceivers }
In relation to ($\ref{eqn_s}$) a PD-NOMA e.g. \cite{IEEElee:nomaser} can be seen as a special case by setting $\delta_{k}=0, \forall k$, i.e. the BS does not adapt transmitted signal constellations. It uses SIC receiver which works on the assumption that users that are closer to the BS achieve higher SNR compared to users that are further away to form easily discernible clusters. Hence a receiver for a near user $l$ can readily decode the data sent to far users $l+1,..,K$, then use the estimated data $\hat{x}_{l+1},..,\hat{x}_{K}$ to regenerate their interference and subtract the sum from the total received signal $r_{l}$, to form a more reliable estimate $\hat{x}_{l}$ as: 
\begin{equation}
\label{eqn_sic}
\hat{x}_{l} = \arg \min_{x_{l}\in \mathbb{X}_{l}}\abs{r_{l}-\sum \limits_{k=l+1}^{K}\hat{h}_{l}\sqrt{\alpha_{k}P_{t}}\hat{x}_{k}-\hat{h}_{l}\sqrt{\alpha_{l}P_{t}}x_{l}}^{2}, 
\end{equation}
where, $\mathbb{X}_{l}=[x_{1},x_{2},..,x_{M_{l}}]^{T}$. SIC work satisfactorily when proper power ratio is used such that $\alpha_{l+1}/\alpha_{l}>>1$ \cite{IEEEassaf:powernoma},\cite{IEEEkara:powqam}. The ratio, if unoptimized can lead to inevitable error floors even if SNR increases monotonically \cite{IEEElee:nomaser}, \cite{IEEEguan:phjd}. Although the entropy maximizing phase rotation method of \cite{IEEEwang:rotsic} show small gain over PD-NOMA with SIC, it too can not resolve error floor trends when power ratio is not satisfied. 

%For the case of $K=2$, assuming user $2$ has worse channel than user $1$, the power allocation rule is given in \cite{IEEElee:nomaser} as follows: 
%
%\begin{equation}
%\label{eqn_sick}
%\alpha_{2} > \frac{(M_{1}-1)(\sqrt{M_{2}f}-1)^{2}}{M_{2}-1+(M_{1}-1)(\sqrt{M_{2}}-1)^{2}},
%\end{equation}
%while $\alpha_{1}=1-\alpha_{2}$.

\vspace{-3mm}

\subsection{JD-NOMA Transceivers with Phase Rotation}
The JD-NOMA \cite{IEEEguan:phjd}, \cite{IEEEchang:phsic} can address some limitations of PD-NOMA with SIC. But their operational power ranges and phase rotations are dictated by separate equations specific for each modulation and power setups to restrict points to I/Q axes to form discernible clusters. Their receivers extract data $\hat{x}_{k}$ by comparing $r_{k}$ against composite signals of all possible transmitted data combinations $\mathbb{X}_{k}, k=1,2,..,K$ as follows: 
\begin{equation}
\label{eqn_jdnoma_recv}
\hat{x}_{k} = \arg \min_{x_{1}\in \mathbb{X}_{1},..x_{K}\in \mathbb{X}_{K}}\abs{r_{k}-\hat{h}_{k}\sum \limits_{k=1}^{K}\sqrt{\alpha_{k}P_{t}}x_{k}}^{2}.
\end{equation}
As ($\ref{eqn_jdnoma_recv}$) cannot resolve detection ambiguity when points cross I/Q axes, phase rotations applied for JD-NOMA are limited to small values. This leads to lower $d^{2}_{min}$ as shown in Figure $\ref{fig_const}$.   
%As noted in \cite{IEEEchang:phsic} JD-NOMA with $K=2$, and $M_{1},M_{2}\in\{16,4\}$ works when $\alpha_{2} \geq 9/14$ only. 
%It is able to separate users' data when constellation points are well separated with good $d^{2}_{min}$. 

\vspace{-3mm}

\subsection{Proposed ACMA Transceivers}
%The real issues of the existing NOMA schemes start to get unraveled when co-channel users have similar SNR and they need to be allocated similar powers i.e. we have to dispense with the degraded broadcast channel model assumed commonly in the research literatures \cite{IEEEhanzo:noma}. As an example, we show in Figure $\ref{fig_const}$, crosses are the constellation points from superposition coding of users in PD-NOMA. In constrast, pluses are from ACMA. When for example $K=2, \alpha_{1}=0.5, \alpha_{2}=0.5, M_{1}=4, M_{2}=4$, we see that only $9$ unique constellation points are visible for PD-NOMA
%%$[-1.0000-1.0000i; -1.0000+0.0000i; 0.0000-1.0000i; 0.0000+0.0000i; -1.0000+0.0000i; -1.0000+1.0000i; 0.0000+0.0000i, 0.0000+1.0000i, 0.0000-1.0000i, 0.0000+0.0000i, 1.0000-1.0000i, 1.0000+0.0000i, 0.0000+0.0000i, 0.0000+1.0000i, 1.0000+0.0000i, 1.0000+1.0000i]$, 
%where there are 4 instances of $0.0000+0.0000j$, 2 instances of $1.0000+0.0000j$, 2 instances of $-1.0000+0.0000j$, 2 instances of $0.0000+1.0000j$, 2 instances of $0.0000-1.0000j$, and the rest 4 single instances. For such signals, there is no way a receiver will be able to seperate all 16 possible transmitted data combinations. In contrast, with ACMA as User 1 signal is phase shifted by $\delta_{1}=\pi/6$ radian; when the users' signals are combined, all 16 constellation points are separated clearly so a receiver can search over them to correctly identify the data.
Continuing with the Figure $\ref{fig_const}$, we can see that phase rotation optimization of the ACMA is not limited to be within the I/Q boundaries i.e. points are allowed to cross over. Moreover they are not limited to $K=2$ and certain modulations only $[M_{1},M_{2}]\in [4,4], [16,4]$ as in \cite{IEEEguan:phjd}, \cite{IEEEchang:phsic}. With ACMA the BS transmits composite data $s$ which uses constellation points by adjusting each user's signal by a phase offset value $\delta_{k}$ given by an algorithm that autonomously adapts it to system parameter changes $K$, $\alpha_{k},M_{k}, k=1,2,..,K$ until the best $d^{2}_{min}$ found. The spread of constellation points becomes progressively denser with increase in $k=1,2,..,K$. But each iteration is independent, adding complexity only linearly. The algorithm is detailed in Table I. 
\begin{table}[!t]
%\fontsize{9}{8}\selectfont
\renewcommand{\arraystretch}{1.4}
\caption{ACMA Algorithm for Calculating Phase Offsets}
\label{table_ACMA_delta}
%\fontsize{9}{9}\selectfont
\centering
\begin{tabular}{l}
\hline
1) Obtain input parameters: 
$K, M_{k}, P_{T}, \alpha_{k}; k=1,2,..,K$.\\
2) Set all possible phase offset matrix for all users of size $K\times V$ to 0,\\
~~~~${\bf{\delta}}={[\bf{0}]}_{K\times V}$.\\
3) Initialize a vector of all $d^{2}_{min}$ values for all users,\\
~~~~${\bf{d}}^{2}_{min(k)}={[\bf{0}]}_{K\times V}$.\\
\hline
%\bfseries First & \bfseries Next\\
%\hline\hline
4) for $k=1:K$ \\
5) ~~~if $k=K$\\
6) ~~~~~$\delta_{k}=\delta_{k,v}$\\
7) ~~~else\\
8) ~~~~~ for $v_{k}=1:V$\\
9)   ~~~~~~~	a) For each $v$, calculate $d^{2}_{\min}$ using ($\ref{eqn_dmin}$); \\
10) ~~~~~~	b) Update $\delta_{k,v}=\frac{2\pi}{V-v+1}$;\\
11) ~~~~~end \\
12) ~~~Pick the $\delta^{k,v}$ that gives maximum $d^{2}_{min}$; \\
13) ~~~Update final $\delta_{k}=\delta_{k,v}$;\\ 
14) ~~~end.\\
15) end.\\	
16) Return $\bf{\delta}$$=[\delta_{1},\delta_{2},..,\delta_{K}]$.\\  	
\hline
\end{tabular}
\end{table} 
%\vspace{-6mm}
\begin{figure}[!t]
\centering
\includegraphics[width=3.5in, height=1.8in]{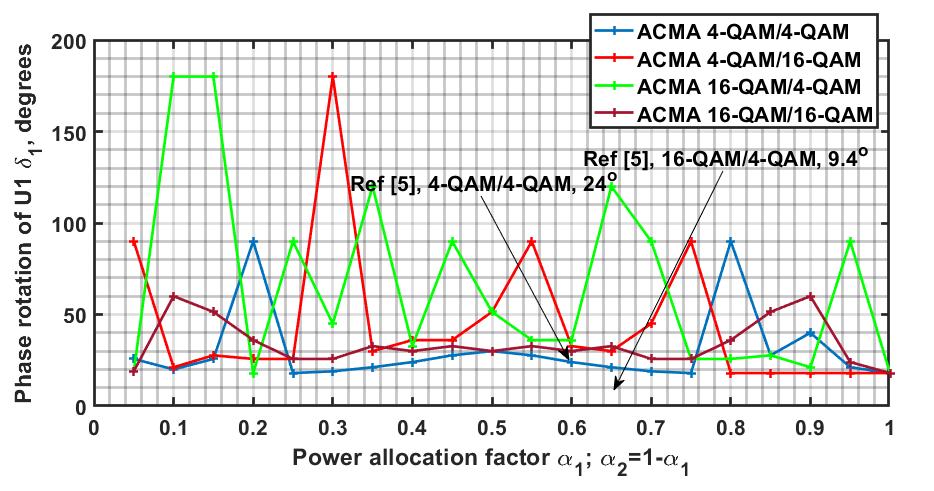}
\caption{Phase rotation $\delta_{1}$ of user U1 for ACMA for $K=2$, different $M_{1}$-QAM$/M_{2}$-QAM configurations, $V=20, Q=20$ under different power sharing values of $\alpha_{1}$  and $\alpha_{2}$. Rotation values from JD-NOMA \cite{IEEEchang:phsic} are also included for reference. }
\label{fig_rota}
\vspace{-6mm}
\end{figure}
           
%\begin{equation}
%\label{eqn_dmin}
%\overline{d}^{2}_{C} = \min\Big\{ d^{2}_{a,b}\Big\}, \forall a, \forall b, a \neq b,
%\end{equation}

%\begin{equation}
%\label{eqn_dmin}
%\begin{split}
%d^{2}_{\min}=\min\Big\{d^{2}_{a,b}\Big\}= \abs{\sum\limits_{k=1}^{K}\sqrt{\alpha_{k}P_{T}}x^{a}_{k}\exp(j\theta_{k}+\delta_{k})-\sum\limits_{k=1}^{K}\sqrt{\alpha_{k}P_{T}}x^{b}_{k}\exp(j\theta_{k}+\delta_{k})}^{2}, a\neq b,\\
%%d^{2}_{\min}=\min\Big\{d^{2}_{a,b}\Big\}= \min{\abs{\sum \limits_{k=1}^{K}\sqrt{\alpha_{k}P_{T}}x_{k} \exp(j\theta_{k}+\delta_{k})}}-\sum \limits_{k=1}^{K}\sqrt{\alpha_{k}P_{T}}x_{k} \exp(j\theta_{k}%+\delta_{k})}^{2}, a\neq b,
%%x^{a}_{k}\in \mathbb{X}= \Big\{[x^{1}_{1},x^{1}_{2},..x^{1}_{K}] [x^{2}_{1},x^{1}_{1},..x^{1}_{K}]... [x^{M_{1}}_{K},x^{M_{2}}_{2},..x^{M_{K}}_{K}]\Big\}; \\
%%x^{b}_{k}\in \mathbb{X}= \Big\{[x^{1}_{1},x^{1}_{2},..x^{1}_{K}] [x^{2}_{1},x^{1}_{1},..x^{1}_{K}]... [x^{M_{1}}_{K},x^{M_{2}}_{2},..x^{M_{K}}_{K}]\Big\}. \\
%%x^{b}_{k}\in \mathbb{X}= \Big\{[x^{1}_{1},x^{2}_{1},..x^{M_{1}}_{1}] [x^{1}_{2},x^{2}_{2},..x^{M_{2}}_{2}]... [x^{1}_{K},x^{2}_{2},..x^{M_{K}}_{K}]\Big\}. \\
%\end{split}
%\end{equation}

Here $d^{2}_{\min}$ is selected from all possible $d^{2}_{a,b}$ that are the distances of a transmitted ACMA signal from all possible $K$ users' data combinations of size $\prod_{k}^{K}M_{k}$ against all other $\prod_{k}^{M}M_{k}-1$ signal possibilities as given in ($\ref{eqn_dmin}$). The $d^{2}_{\min}$ values are then calculated for $V$ phase offset values $\delta_{k,v}, v=1,2,..,V$ covering the full $2\pi$ radians and the one that gives highest $d^{2}_{\min}$ is updated to give $\delta_{k}$.  Note that full set of $\delta_{k}, k=1,2,..,K$ values for all configurations $P_{T}, M_{k},\alpha_k, \forall k$ $Q=P_{T}/\min(\alpha_{k}-\alpha_{l}), k\neq l$, can be calculated offline and stored in memory as lookup tables, its use does not add latency. We show a plot of these values in Figure $\ref{fig_rota}$ for $M_{k}\in\{4,16\}, \forall k$, under the full range of $[\alpha_{1},\alpha_{2}]$. The complete ACMA transceiver algorithm is given Table II.

The system can also be configured to operate in both static and dynamic user environments. For the former, the BS sets $\phi_{k}, k=1,2,..,K$ only once after all receivers have been calibrated. For the latter, the BS adjusts paramaters including $\delta_{k}, k=1,2,..,K$ upon the signal quality feedback from receivers. The BS shares info $K, P_{T}, M_{k}, \alpha_{k}, k=1,2,..,K$ to receivers. This may be done via downlink control information (DCI) slots of PDCCH of LTE, 5G NR standards. Signalling overheads for this are expected be small. 

\begin{figure*}[!t]
% ensure that we have normalsize text
\normalsize
% Store the current equation number.
\setcounter{MYtempeqncnt}{\value{equation}}
% Set the equation number to one less than the one
% desired for the first equation here.
% The value here will have to changed if equations
% are added or removed prior to the place these
% equations are referenced in the main text.
\setcounter{equation}{3}
\begin{equation}
%\label{eqn_dbl_y}
%pe_{C} =\frac{\sqrt{M_{\Sigma}}-1}{M_{\Sigma}}\Bigg\{ (\sqrt{M_{\Sigma}}-1)+4{ {1-\mu_{C}} \choose{2}}- (\sqrt{M_{\Sigma}}-1)\Bigg\{\frac{4}{\pi}\mu_{C}\tan^{-1} (\mu_{C}) \Bigg\}\Bigg\}
%\frac{p_{k}\abs{{\sum\limits_{m\in\bf{\iota}_{k}} }g^{*}_{k,m}g_{k,m}}^{2}}   {\underbrace{\sum \limits_{i=1, i\neq k}^{K}p_{i}\abs{{\sum\limits_{m\in\bf{\iota}_{k}} } g^{*}_{k,m} g_{i,m}s_{i}-g^{*}_{k,m}g_{i,m}\hat{s}^{l-1}_{i}}^{2}}_{\mathrm{IC}}+\underbrace{\sum \limits_{k=1, i=1}^{K}p_{k}\abs{{\sum\limits_{m\in\bf{\iota}_{k}} } \{\hat{g}_{k,m}-g_{k,m}\}\{\hat{g}_{i,m}-g_{i,m}\}}^{2}}_{\mathrm{RMUI}}+\sigma^{2}_{z} {\sum\limits_{m\in\bf{\iota}_{k}} } \abs{\hat{g}_{k,m}}^{2}}.
\label{eqn_dmin}
d^{2}_{\min}=\min\Big\{d^{2}_{a,b}\Big\}= \arg \min_{\delta_{k,v}\in \mathbb{\delta}}\abs{\sum\limits_{k=1}^{K}\sqrt{\alpha_{k}P_{T}}x^{(a)}_{k}\exp(j\theta_{k}+\delta_{k,v})-\sum\limits_{k=1}^{K}\sqrt{\alpha_{k}P_{T}}x^{(b)}_{k}\exp(j\theta_{k}+\delta_{k,v})}^{2}, \forall a\neq b.
\end{equation}
% Restore the current equation number.
%\setcounter{equation}{\value{MYtempeqncnt}}
% The IEEE uses as a separator
%\hrulefill
% The spacer can be tweaked to stop underfull vboxes.
\vspace*{-5mm}
\end{figure*}

\begin{figure*}[!t]
% ensure that we have normalsize text
\normalsize
% Store the current equation number.
\setcounter{MYtempeqncnt}{\value{equation}}
% Set the equation number to one less than the one
% desired for the first equation here.
% The value here will have to changed if equations
% are added or removed prior to the place these
% equations are referenced in the main text.
\setcounter{equation}{4}
\begin{equation}
%\label{eqn_dbl_y}
%pe_{C} =\frac{\sqrt{M_{\Sigma}}-1}{M_{\Sigma}}\Bigg\{ (\sqrt{M_{\Sigma}}-1)+4{ {1-\mu_{C}} \choose{2}}- (\sqrt{M_{\Sigma}}-1)\Bigg\{\frac{4}{\pi}\mu_{C}\tan^{-1} (\mu_{C}) \Bigg\}\Bigg\}
%\frac{p_{k}\abs{{\sum\limits_{m\in\bf{\iota}_{k}} }g^{*}_{k,m}g_{k,m}}^{2}}   {\underbrace{\sum \limits_{i=1, i\neq k}^{K}p_{i}\abs{{\sum\limits_{m\in\bf{\iota}_{k}} } g^{*}_{k,m} g_{i,m}s_{i}-g^{*}_{k,m}g_{i,m}\hat{s}^{l-1}_{i}}^{2}}_{\mathrm{IC}}+\underbrace{\sum \limits_{k=1, i=1}^{K}p_{k}\abs{{\sum\limits_{m\in\bf{\iota}_{k}} } \{\hat{g}_{k,m}-g_{k,m}\}\{\hat{g}_{i,m}-g_{i,m}\}}^{2}}_{\mathrm{RMUI}}+\sigma^{2}_{z} {\sum\limits_{m\in\bf{\iota}_{k}} } \abs{\hat{g}_{k,m}}^{2}}.
\label{eqn_acma_recv}
\hat{\bf{x}} = \arg \min_{x_{1}\in \mathbb{X}_{1},..x_{K}\in \mathbb{X}_{K} \hat{\delta}_{1}\in {\bf{\delta}},\hat{\delta}_{2}\in {\bf{\delta}},..,\hat{\delta}_{K}\in {\bf{\delta}}}\abs{r_{k}-\hat{h}_{k}\sum \limits_{k=1}^{K}\sqrt{\alpha_{k}P_{T}}x_{k} \exp(j\theta_{k}+\hat{\delta}_{k})}^{2}.
\end{equation}
% Restore the current equation number.
%\setcounter{equation}{\value{MYtempeqncnt}}
% The IEEE uses as a separator
\hrulefill
% The spacer can be tweaked to stop underfull vboxes.
\vspace*{-5mm}
\end{figure*}

The receiver estimates transmitted data by using the MML joint detection method given in ($\ref{eqn_acma_recv}$), which compares $r_{k}$ against the reconstructed signal of all likely data $x_{k}\in \mathbb{X}_{K}, k=1,2,..,K$, $\hat{\delta}_{k}$ and shared parameters $P_{T},K, \alpha_{k}, M_{k}, k=1,2,..,K$. It also estimates all possible $\bf{\hat{\delta}}$ independently using the algorithm in Table I and store in memory. The process in ($\ref{eqn_acma_recv}$) uses a simple lookup process for picking phases $\hat{\delta}_{k},k=1,2,..,K$ which does not introduce extra latency.  
%\begin{equation}
%\label{eqn_acma_recv}
%\hat{x}_{k} = \arg \min_{x_{1}\in \mathbb{X}_{1},..x_{K}\in \mathbb{X}_{K}, \delta^{k}_{v}\in \bf{\delta}}\abs{r_{k}-\hat{h}_{k}\sum \limits_{i=1}^{K}\sqrt{\alpha_{i}P_{t}}x_{i}}^{2}.
%\end{equation}

%including $d^{2}_{min}$ optimised phase offset values $\delta_{k}, k=1,2,..,K$

\begin{table}[!t]
\renewcommand{\arraystretch}{1.3}
\caption{The ACMA Transceiver Algorithm Steps}
\label{table_ACMA}
\centering
\begin{tabular}{l}
\hline
\bf{Basestation Transmitter}\\
\hline
%\bfseries First & \bfseries Next\\
%\hline\hline
1) Obtain input parameters: \\
$P_{T}$, $M_{k}$, $\alpha_{k}$, $\Psi, \phi^{k}, k=1,2,..,K$.\\
2) Map each user's data bits into a M-QAM/M-PSK signal $x_{k}$.\\ 
if $\Psi=0$\\
\emph{\underline{Dynamic users' environment}}\\
3) ~~~Update $\theta^{k}=\angle(x_{k}), \forall k$.\\
4) ~~~Calculate $\delta^{k}, k=1,2,..,K$ using the algorithm in Table I.\\
else\\
\emph{\underline{Static users' environment}}\\
5) ~~~Set $\theta^{k}=\angle(x_{k})$; $\delta_{k}=\phi^{k}, k=1,2,..,K$.\\
end
\\
6) Form composite ACMA signal as per ($\ref{eqn_s}$).\\
7) Share $P_{T}, K, V, Q$ $\alpha_{k}, M_{k} \forall k$ to users via control channels.\\
8) Transmit signal $s$.\\
\hline
\bf{Receiver $k^{th}$ user}\\
\hline
1) Decode the shared data $P_{T}, K, V, Q$ $\alpha_{k}, M_{k}, \forall k$.\\
2) Calculate $\bf{\hat{\delta}}$ using algorithm in Table I and store as lookup.\\
3) Initialize users' likely transmitted signals $\hat{\bf{x}}=[0,0,..,0]_{K\times 1}$.\\
4) Obtain $\hat{\delta}_{k},\forall k$ for each $\alpha_{k}, M_{k} \forall k$ update using the lookup table.\\
5) Update $\hat{\bf{x}}$ using the MML detection method in ($\ref{eqn_acma_recv}$).\\
6) Extract desired data $\hat{x}_{k}\in\hat{\bf{x}}$.\\
7) Report estimates of received power from $r_{k}$ to BS periodically.\\
\hline
\end{tabular}
\end{table}

%Note that the receivers can recover transmitted data $\hat{x}_{k},\forall k$ even if BS does not share $\delta_{k}$ values if it knows $V$ by using ($\ref{eqn_acma_recv}$) for all possible phase offsets.
 
\vspace{-4mm}

\subsection{Comparison of Computational Efforts}
In Table III we compare computational efforts of ACMA transceivers against PD-NOMA, JD-NOMA. We simplify expressions using an average of users' modulation alphabets $\mu=\sum_{k=1}^{K}M_{k}/K$. PD-NOMA requires lower computational efforts than ACMA and JD-NOMA. This is mainly due the latter using joint detection of all $K$ users signals at the receiver side scaling at the rate of $\mu^{K}$ searches. ACMA requires some additional offline pre-processing that needs $V$ phase offset searches against $Q$ possible power steps per user to arrive at the $d^{2}_{min}$ optimized constellations, totalling $KQV(\mu K)^{2}$ multiplications. It should be noted that number of co-channel users $K$ for any NOMA access is essentially small e.g. $2$, $3$ in practice which makes computational demand less of an issue for online ACMA data detection algorithms.     
\begin{table}[!t]
\caption{Computational Efforts Of ACMA Transceivers Against PD-NOMA and JD-NOMA}
%\label{table_example}
\centering
%% Some packages, such as MDW tools, offer better commands for making tables
%% than the plain LaTeX2e tabular which is used here.
\begin{tabular}{c|c|c|c}
\hline
{Scheme} & {Multiply}& (Add)/(Subtract) & Offline Process\\
\hline
\hline
PD-NOMA \cite{IEEElee:nomaser} & $K\mu+K\mu$ &$(K/2)/(K^{2}/2)$& $-$\\
\hline
JD-NOMA \cite{IEEEguan:phjd}, \cite{IEEEchang:phsic} & $K\mu+\mu^{K}$ &$(K)/(K)$& $K^{2}Q\mu\log(V)$\\
\hline
ACMA & $K\mu+\mu^{K}$ &$(2K)/(K)$& $KQV(\mu K)^{2}$\\
\hline
\end{tabular}
\vspace{-4mm}
\end{table}

\vspace{-1mm}
\section{Numerical Results and Discussions}
Here we use simulation results to compare symbol error rate (SER) and throughput of ACMA against the other schemes. %For fair comparisons, JD-NOMA \cite{IEEEguan:phjd}, \cite{IEEEchang:phsic} with the proposed MML detection ($\ref{eqn_acma_recv}$) also used where possible. 
Two-user system $K=2$ used for clarity and ease of comparisons, with M-QAM signals $M_{k}\in \{4,16\}$ and vary $\alpha_{1}\in [0-1]$ where $\alpha_{2}=1-\alpha_{1}$ for all schemes. Systems using $K>2$ add extra computational efforts, but the performace trends of all schemes are expected to continue. For ACMA $Q=20, V=20$ and $\Psi=0$ assumed. Estimation of channel and other parameters is assumed perfect $\hat{h}_{k}=h_{k}, \hat{\delta}_{k}=\delta_{k}, \forall k$. In practice carrier frequency, channel and phase are estimated using phase locked loop and reference symbols. Estimation error impact SER and specially when $M_{k}$ is high \cite{GoldS:WCom}.      

\begin{figure}[!t]
\centering
\includegraphics[width=3.2in, height=2.6in]{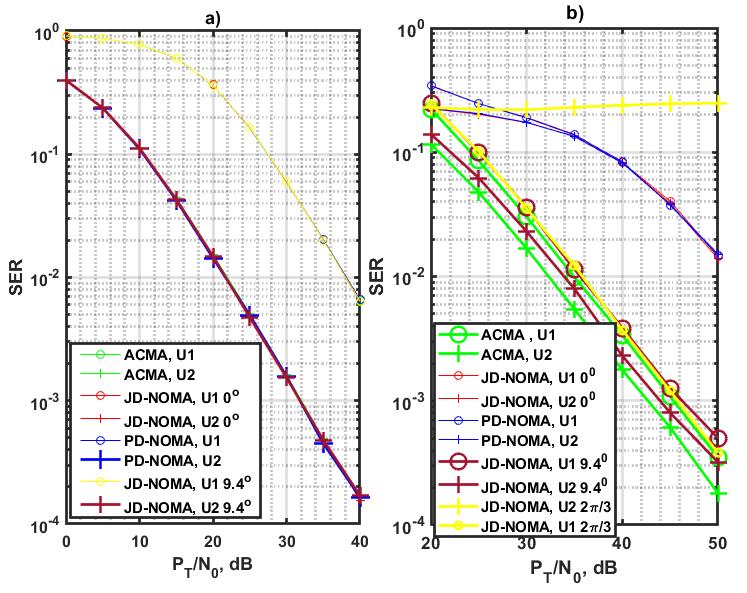}
\caption{SER of users U1 and U2 for ACMA compared with JD-NOMA, PD-NOMA for $K=2$, $\abs{h_{1}}^{2}/\abs{h_{2}}^{2}=0$ dB, $M_{1}=16$, $M_{2}=4$ and for $[\alpha_{1},\alpha_{2}]=$ a) $[0.1,0.9]$, b) $[0.35, 0.65]$.}
\label{fig_sim4}
\vspace{-6mm}
\end{figure}

In Figure $\ref{fig_sim4}$ a) and b), we present SER of ACMA for $K=2$ and $M_{1}=16, M_{2}=4$ under the same total SNR, $P_{t}/N_{0}$ and how they compare against other NOMA schemes under different [$\alpha_{1}$, $\alpha_{2}$]  values of $[0.1, 0.9]$ and $ [0.35,0.65]$, respectively. The users are labeled as U1 and U2. The SER curves for all schemes are identical for $[\alpha_{1},\alpha_{2}]=[0.1,0.9]$. This is expected because under controlled power allocation the SIC method in (\ref{eqn_sic}) is able to successfully decode the signal of U1, subtract its interference to U2 to obtain cleaner signal giving the same results as the joint detection method \cite{IEEEassaf:powernoma}-\cite{IEEEkara:powqam}. 
%The JD-NOMA achieve identical SERs of NOMA SIC as composite signals with $[\alpha_{1},\alpha_{2}]=[0.1,0.9]$ still gives 4-QAM signal with constellation points fully separated across the $4$ quadrants. The SER of U2, %which is allocated more power is better than U1 as expected. 
%The JD-NOMA SER upper bounds obtained from (\ref{eqn_serc}) are plotted to show how tight these upper bounds are. The $Pe_{C}$ curve for $\alpha_{1}/\alpha_{2}=0.17/0.83$ matches the simulated SER very well, %indicating that distances are evenly distributed around $\overline{d}^{2}_{C}$. The $Pe_{C}$ for $\alpha_{1}/\alpha_{2}=0.3/0.7$ are higher than that of simulated SERs. This can be attributed to the uneven %distibution of distances with more points giving distances greater than $\overline{d}^{2}_{C}$. 
It is worth noting that the SER of ACMA depends on how closely spaced all the possible composite signals of users' are relative to each other regardless of which quadrant the constellation points are located i.e. $d^{2}_{min}$ in ($\ref{eqn_dmin}$). For $[\alpha_{1},\alpha_{2}]=[0.35, 0.65]$ ACMA offers better SER compared with others. It outperforms JD-NOMA even with its optimal rotation of $9.4^{\circ}$ by $\approx 2$ dB at the SER of $10^{-2}$. This is because it achieves greater $d^{2}_{min}$ by adapting phase offset of U1 to reach highest $d^{2}_{min}$ with $\delta_{1}=2\pi/3$. PD-NOMA and JD-NOMA without phase rotation ($0^{\circ}$) lag ACMA by $> 15$ dB due to increase in unresolved magnitude and phase noise. For completeness we also plot SER of JD-NOMA with rotation of $2\pi/3$ to U1 signal. As can be seen from Figure 3 b), while U2 performs close to ACMA, U1 suffers error floor as the JD-NOMA receiver has no way to resolve it by identifying the detection boundaries from the composite signals. 
%Figure 3 shows the SER of JD-NOMA for $K=2$ and $M_{1}=16, M_{2}=16$, with two users labeled as U1 and U2, respectively for different $\alpha_{1}$, $\alpha_{2}$ values and how they compare against NOMA under the same total SNR, $P_{t}/N_{0}$. The SER upper bounds calculated using (\ref {eqn_serc}) are also plotted. Similar conclusions to that in Figure 2 can be made here. The bound $Pe_{C}$ for $\alpha_{1}/\alpha_{2}=0.15/0.85$ show higher SER than simulated SER but the gap is less. This can be attributed to higher concentration of $d^{2}_{a,b}$ that are closer to $\overline{d}^{2}_{C}$ than in the previous case.
%
%
%
%\begin{figure}[!t]
%\centering
%\includegraphics[width=3.5in, height=2.8in]{ACMA vs JD-NOMA 16-16QAM 0.5-0.5 and single user 256-MQAM SER.jpg}
%\caption{SER of users U1 and U2 for JD-NOMA compared with NOMA for $K=2$, for $M_{1}=16$, $M_{2}=4$ and different $\alpha_{1}$, $\alpha_{2}$ values.}
%\label{fig_sim}
%\vspace{-4mm}
%\end{figure}

\begin{figure}[!t]
\centering
\includegraphics[width=3.2in, height=2.6in]{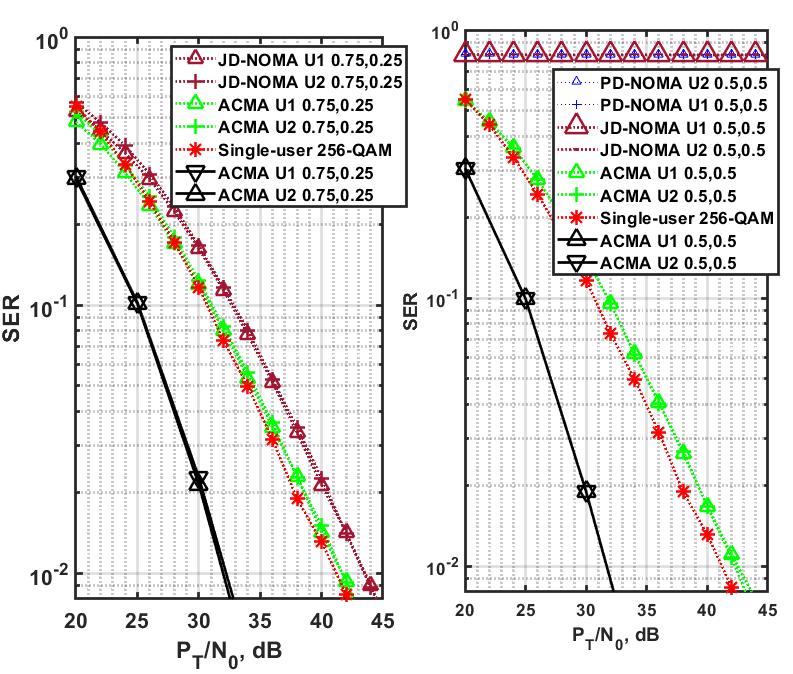}
\caption{SER of U1 and U2 for ACMA  compared with JD-NOMA, PD-NOMA for $K=2$, $M_{1}=16$, $M_{2}=16$, $\abs{h_{1}}^{2}/\abs{h_{2}}^{2}=0$ dB, $[\alpha_{1}, \alpha_{2}]=$ a) $[0.75,0.25]$ and b) $ [0.5,0.5]$. Two-antenna receive diversity (Solid lines).}
\label{fig_sim5}
\vspace{-6mm}
\end{figure}

In Figure $\ref{fig_sim5}$ a) and b), we show SER with $M_{1}=16, M_{2}=16$ for different $[\alpha_{1}, \alpha_{2}]$ of $[0.75, 0.25]$ and $ [0.5,0.5]$, respectively. Under higher order $M_{k}$, constellation points become closer with much higher probabilities of them crossing the boundaries of users' modulated signals. The JD-NOMA schemes in \cite{IEEEguan:phjd}, \cite{IEEEchang:phsic} do not provide phase rotation parameters for $M_{1}=16, M_{2}=16$. Looking at Figure 1 we expect phase rotation to deminish $\rightarrow 0$. ACMA achieves $\approx 2$ dB gain over JD-NOMA at the SER of $10^{-2}$ for $[\alpha_{1},\alpha_{2}= [0.75,0.25]$ as it finds better $d^{2}_{min}$ by adjusting $\delta_{1}$. For $[\alpha_{1},\alpha_{2}]= [0.5, 0.5]$, JD-NOMA exhibits error floor similar to PD-NOMA, but ACMA does not. Even under such an extremely tight constellation signals, ACMA achieves SER of $10^{-2}$ at the cost of  $\approx  2$ dB over single-user $256$-QAM. We also use simple analysis to estimate its SER of ACMA. As it does not treat co-channel users as noise or interference, we can use the SNR vs. SER relation of ordinary M-QAM signals of size  $\Sigma=\prod_{k=1}^{K}M_{k}$ and adjust it by a ratio $\min\{d^{2}_{ACMA}\}/\min\{d^{2}_{\Sigma-QAM}\}$ to give its upper bound: $SER_{ACMA} \leq {SER}_{\Sigma-QAM} \Big\{P_{T}/N_{0} \times \min\{d^{2}_{ACMA}\}/\min\{d^{2}_{\Sigma-QAM}\}\Big\}$. This approach predicts a max SER slope of ACMA $2.6$ dB worse than that of single-user $256$-QAM--which the results in Figure $\ref{fig_sim5}$ support. It is clear that, for higher $K$ higher SNR is required to decode users' data. Use of spatial diversity can significantly reduce the SNR required in such cases. Figure $\ref{fig_sim5}$ shows SER gains of $\approx 10$ dB with two receive antennas for both cases. 
%Similar conclusions to that in Figure 2 can be made here. The bound $Pe_{C}$ for $\alpha_{1}/\alpha_{2}=0.15/0.85$ show higher SER than simulated SER but the gap is less. This can be attributed to higher concentration of $d^{2}_{a,b}$ that are closer to $\overline{d}^{2}_{C}$ than in the previous case.

\begin{figure}[!t]
\centering
\includegraphics[width=3.2in, height=2.6in]{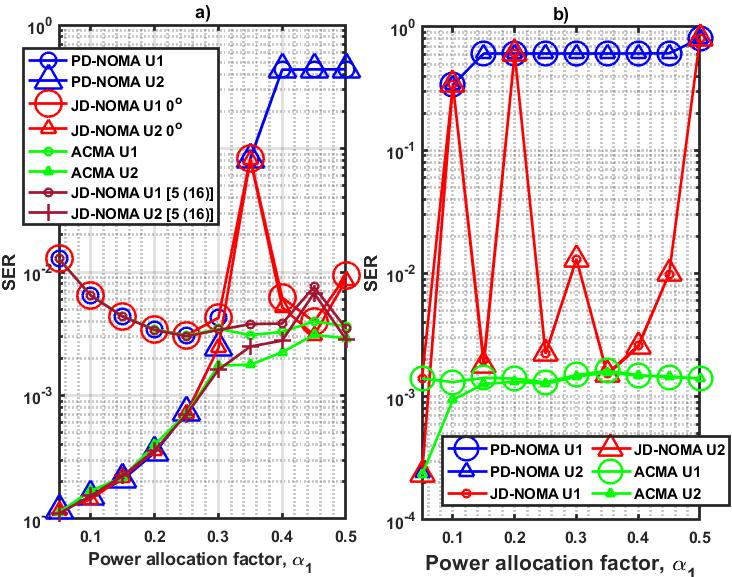}
\caption{SER of ACMA for U1 and U2 compared with JD-NOMA and PD-NOMA under $\abs{h_{1}}^{2}/\abs{h_{2}}^{2}=0$ dB for a) $M_{1}=16, M_{2}=4, P_{t}/N_{0}=40$ dB, and  b) $M_{1}=16, M_{2}=16, P_{t}/N_{0}=50$ dB, for different $\alpha_{1}$ values in the range $[0.05-0.5]$ where $\alpha_{2}=1-\alpha_{1}$.}
\label{fig_sim6}
\vspace{-4mm}
\end{figure}

In Figure $\ref{fig_sim6}$ we compare SER of ACMA for $M_{1}=16, M_{2}=4$, and $M_{1}=16, M_{2}=16$, while varying $\alpha_{1}$, ($\alpha_{2}=1-\alpha_{1}$) until they use equal powers of $0.5$ for the same $P_{t}/N_{0}$ dB. In the Figure $\ref{fig_sim6}$ a), PD-NOMA shows degraded SER for $\alpha_{1}>0.1$. JD-NOMA uses optimal rotation of (16) in \cite{IEEEchang:phsic}, giving improved SER compared to PD-NOMA. Both schemes fall short of SER of ACMA when powers allocated to users are similar i.e $\alpha_{1}\rightarrow 0.5$. In Figure $\ref{fig_sim6}$ b), for SER trends continue. Here we use JD-NOMA with $0^{\circ}$ phase rotation for U1. ACMA continues to show better SER, specially when $\alpha_{1}\rightarrow 0.5$. The resilience of ACMA can make it a very attractive choice to ease grouping of users without needing them to be at certain distance ratios from the BS \cite{IEEEzhang:reliablenoma} and to sustain data communications in extreme conditions. 

\begin{figure}[!t]
\centering
\includegraphics[width=3.5in, height=1.7in]{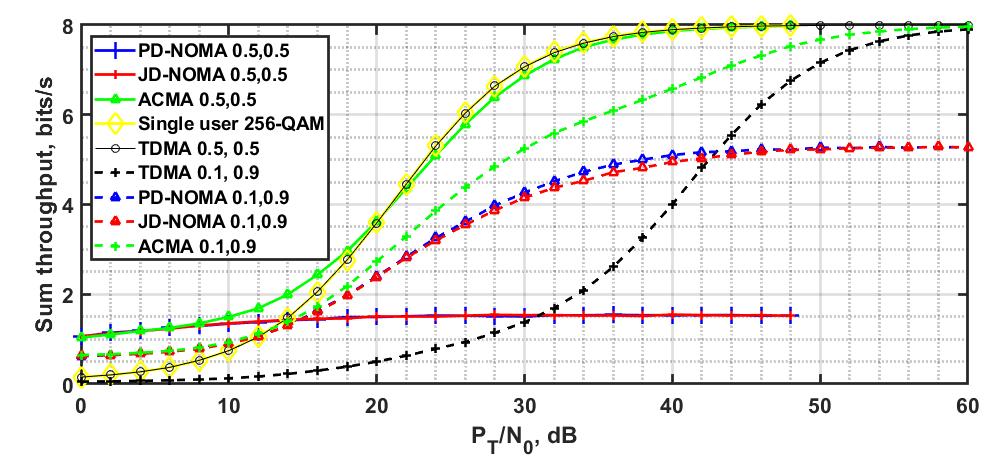}
\caption{Sum throughput of ACMA compared with TDMA, JD-NOMA and PD-NOMA for $K=2$ $[M_{1},M_{2}] = [16,16], [\alpha_{1},\alpha_{2}]=[0.1,0.9] $ and $[0.5,0.5] $. Solid lines are for $\abs{h_{1}}^{2}/\abs{h_{2}}^{2}=0$ dB and dotted lines for $\abs{h_{1}}^{2}/\abs{h_{2}}^{2}=20$ dB.}
\label{fig_sim7}
\vspace{-6mm}
\end{figure}

Continuing for $K=2, M_{1}=16, M_{2}=16$, in Figure $\ref{fig_sim7}$ we compare sum data throughputs $\Omega$ of ACMA for different $[\alpha_{1},\alpha_{2}]$ values and for two channel conditions $\abs{h_{1}}^{2}/\abs{h_{2}}^{2}=[0, 20]$ dB. The $\Omega=\sum_{k=1}^{K}\log_{2}(M_{k})[1-$SER$_k]$ is a useful metric in comparing SER achieved to maintain a fixed data rate. We also show SER of TDMA under the same conditions where we allocate proportional power $\alpha_{k}P_{T}$ and time-slot $\alpha_{k}$ resources to users. For $[\alpha_{1},\alpha_{2}]=[0.5,0.5], \abs{h_{1}}^{2}/\abs{h_{2}}^{2}=0$ dB, ACMA achieves higher throughput compared with JD-NOMA and PD-NOMA. Throughputs of TDMA and single-user $256$-QAM are also plotted, where ACMA shows comparable performance. For $[\alpha_{1},\alpha_{2}]=[0.1,0.9] , \abs{h_{1}}^{2}/\abs{h_{2}}^{2}=20$ dB, ACMA achieves higher sum throughput compared with all other schemes. For example, it achieves $6.5$ bits/s at the $P_{T}/N_{0}=40$ dB compared with $\approx 5$ bits/s for PD-NOMA/JD-NOMA and $4$ bits/s for TDMA. This is because ACMA balances the constellation points to give best possible detection even when they are very tightly spaced for higher $M_{k}$. 
%The results also suggest that under low SNR conditions, it is better to split the power amongst users and use the same time/frequency resource, favoring the non-orthogonal schemes.  

\vspace{-1mm}
\section{Conclusions}
We investigated a new ACMA approach that resolves key limitations of existing NOMA transceivers in practical wireless environments where high channel gain disparity assumption of users cannot always be satisfied. We showed that the PD-NOMA and JD-NOMA can suffer from inevitable error floor if power allocation to grouped users are not set within the required ratios. Such constraints can lead to waste of power or data throughputs due to part of BS power not being utilized fully. We showed that adaptive phase offset approach of ACMA gives flexibility in power, modulation allocations and better minimum distances leading to better SER and data rates. For example, it achieves $6.5$ bits/s compared with $4$ for TDMA and requires $\approx 2$ dB less power than JD-NOMA even with best phase rotation configurations. It shows higher gains especially when users need to be allocated similar powers, e.g. when they are within similar distances from the BS and likely in practice. Computational efforts of ACMA are reasonable for most practical cases. The use of multiple antenna transmission and reception or MIMO approach can increase the throughput and reliability of ACMA much further. Also, investigation of techniques lowering feedback overheads e.g. \cite{IEEEkara:bima} will be interesting for future work.            

\vspace{-1mm}

%given the gains, may be well justified

% or


%\appendices

%%\section{}
%%Appendix two text goes here.

%%\section*{Acknowledgment}

%%The authors would like to thank...

\ifCLASSOPTIONcaptionsoff
  \newpage
\fi


\begin{thebibliography}{1}
\bibitem{IEEElee:nomaser}
I. Lee and J. Kim, “Average Symbol Error Rate Analysis for Non-Orthogonal Multiple Access with M-Ary QAM Signals in Rayleigh Fading Channels,” \emph{IEEE Commun. Letters}, vol. 23, no. 8, pp. 1328–1331,
Jun. 2019.
\bibitem{IEEEmakki:noma}
B. Makki, K. Chitti, A. Behravan and M. -S. Alouini, “A Survey of NOMA: Current Status and Open Research Challenges," \emph{IEEE Open J. Commun. Soc.}, vol. 1, pp. 179-189, 2020.
\bibitem{IEEEguan:phsic}
X. Guan, Q. Yang, Y. Hong, and C. -K. Chan “Non-orthogonal multiple access with phase predistortion in visible light communication,'' \emph{Optics Express}, Vol. 24, No. 22, 31 Oct 2016.
\bibitem{IEEEguan:phjd}
X. Guan, Q. Yang and C. -K. Chan, "Joint Detection of Visible Light Communication Signals Under Non-Orthogonal Multiple Access," \emph{IEEE Photo. Tech. Letters}, vol. 29, no. 4, pp. 377-380, 15 Feb.15, 2017.
\bibitem{IEEEchang:phsic}
Y. Chang and K. Fukawa, ‘Non-orthogonal multiple access with phase rotation employing joint MUD and SIC’, \emph{IEEE Veh. Technol. Conf.}, vol. 2018-June, pp. 1–5, 2018.
\bibitem{IEEEqiu:aljsic}
M. Qiu, Y.-C. Huang, and J. Yuan, ‘Downlink Non-Orthogonal Multiple Access Without SIC for Block Fading Channels: An Algebraic Rotation Approach’, \emph{IEEE Trans. Wirel. Commun.}, vol. 18, no. 8, pp. 3903–3918, 2019.
\bibitem{IEEEwang:rotsic}
N. Ye, A. Wang, X. Li, W. Liu, X. Hou, and H. Yu, ‘On constellation rotation of NOMA with SIC receiver’, \emph{IEEE Commun. Lett.}, vol. 22, no. 3, pp. 514–517, 2018.
\bibitem{IEEElin:rotmd}
C. H. Lin, S. L. Shieh, T. C. Chi, and P. N. Chen, ‘Optimal inter-constellation rotation based on minimum distance criterion for uplink NOMA’, \emph{IEEE Trans. Veh. Technol.}, vol. 68, no. 1, pp. 525–539, 2019.
\bibitem{IEEEng:pmpe}
B. K. Ng and C. T. Lam, ‘Joint Power and Modulation Optimization in Two-User Non-Orthogonal Multiple Access Channels: A Minimum Error Probability Approach’, \emph{IEEE Trans. Veh. Technol.}, vol. 67, no. 11, pp. 10693–10703, 2018.
\bibitem{IEEEzhang:reliablenoma}
Y. Zhang, J. Wang, L. Zhang, Y. Zhang, Q. Li and K. -C. Chen, "Reliable Transmission for NOMA Systems With Randomly Deployed Receivers," \emph{IEEE Trans. on Commun.}, vol. 71, no. 2, pp. 1179-1192, Feb. 2023.
%\bibitem{IEEEzhu:powernoma}
%J. Zhu, J. Wang, Y. Huang, S. He, X. You and L. Yang, "On Optimal Power Allocation for Downlink Non-Orthogonal Multiple Access Systems," \emph{IEEE Jour. on Select. Areas in Commun.}, vol. 35, no. 12, pp. 2744-2757, Dec. 2017, doi: 10.1109/JSAC.2017.2725618.
\bibitem{IEEEassaf:powernoma} 
T. Assaf, A. Al-Dweik, M. S. E. Moursi, H. Zeineldin and M. Al-Jarrah, "NOMA Receiver Design for Delay-Sensitive Systems," \emph{IEEE Systems Journ.}, vol. 15, no. 4, pp. 5606-5617, Dec. 2021
\bibitem{IEEEkara:powqam}
F. KARA and H. KAYA, "A True Power Allocation Constraint for Non-Orthogonal Multiple Access with M-QAM Signalling," \emph{2020 IEEE Micr. Theo. and Techn. in Wireless Commun. (MTTW)}, Riga, Latvia, 2020, pp. 7-12
\bibitem{IEEEkara:bima}
F. Kara, H. Kaya, H. Yanikomeroglu, B. K. Ng and C. -T. Lam, "Bit-Interleaved Multiple Access: Improved Fairness, Reliability, and Latency for Massive IoT Networks," \emph{IEEE Internet of Things Journ.}, vol. 10, no. 18, pp. 16006-16027, 15 Sept.15, 2023
%\bibitem{IEEEwei:constr}
%Z. Wei et al., "Beyond Nonorthogonal Multiple Access: New Role of Constructive Interference," \emph{IEEE Wireless Commun. Letters}, vol. 11, no. 10, pp. 2225-2229, Oct. 2022.
%\bibitem{IEEEassa:nomaqam}
%T. Assaf, A. J. Al-Dweik, M. S. E. Moursi, H. Zeineldin and M. Al-Jarrah, “Exact Bit Error-Rate Analysis of Two-User NOMA Using QAM With Arbitrary Modulation Orders,” \emph{IEEE Commun. Letters}, vol. 24, no. 12, pp. 2705-2709, Dec. 2020.
%\bibitem{IEEEding:noma}
%Z. Ding, R. Schober and H. V. Poor, “Unveiling the Importance of SIC in NOMA Systems—Part 1: State of the Art and Recent Findings," \emph{IEEE Commun. Letters}, vol. 24, no. 11, pp. 2373-2377, Nov. 2020. 
%\bibitem{IEEEding:pairing}
%Z. Ding, P. Fan and H. V. Poor, "Impact of User Pairing on 5G Nonorthogonal Multiple-Access Downlink Transmissions," \emph{IEEE Trans. on Veh. Tech.}, vol. 65, no. 8, pp. 6010-6023, Aug. 2016.
%\bibitem{IEEEpair:noma}
%N. S. Mouni, A. Kumar and P. K. Upadhyay, "Adaptive User Pairing for NOMA Systems With Imperfect SIC," in IEEE Wireless Communications Letters, vol. 10, no. 7, pp. 1547-1551, July 2021, doi: 10.1109/LWC.2021.3074036.
%\bibitem{IEEEdohl:jdnoma}
%M. B. Shahab, S. J. Johnson, M. Shirvanimoghaddam and M. Dohler, "Receiver Design for Uplink Power Domain NOMA With Discontinuous Transmissions," \emph{ IEEE Comm. Letters}, vol. 25, no. 8, pp. 2738-2742, Aug. 2021, doi: 10.1109/LCOMM.2021.3077609.
%\bibitem{IEEEsaito:noma}
%Y. Saito et al., “Non-Orthogonal Multiple Access (NOMA) for Future Radio Access," \emph{Proc. IEEE VTCSpring ’13}, June 2013, pp. 1–5.
%\bibitem{IEEEhanzo:noma}
%Y. Liu, Z. Qin, M. Elkashlan, Z. Ding, A. Nallanathan and L. Hanzo, "Nonorthogonal Multiple Access for 5G and Beyond," \emph{Proceed. of the IEEE}, vol. 105, no. 12, pp. 2347-2381, Dec. 2017.
%\bibitem{IEEETaricco:powernoma}
%G. Taricco, "Fair Power Allocation Policies for Power-Domain Non-Orthogonal Multiple Access Transmission With Complete or Limited Successive Interference Cancellation," \emph{IEEE Access}, vol. 11, pp. 46793-46803, 2023.
%\bibitem{IEEEchoy:powernoma}
%C. Choy Chai and X. -P. Zhang, "Iterative Water-Filling Power and Subcarrier Allocation for Multicarrier NOMA Downlink," \emph{2023 IEEE Intern. Conf. on Acou., Sp. and Signal Proc. (ICASSP)}, Rhodes Island, Greece, 2023, pp. 1-5.
%\bibitem{IEEEniko:scma}
%H. Nikopour et al., "SCMA for downlink multiple access of 5G wireless networks," \emph{2014 IEEE Global Communications Conference}, Austin, TX, USA, 2014, pp. 3940-3945, 2014.
%\bibitem{EURAmao:rsma}
%Y. Mao, B. Clerckx, and V.O.K. Li, “Rate-splitting multiple access for downlink communication systems: bridging, generalizing and outperforming SDMA and NOMA,” EURASIP J. Wireless Commun. Netw., May 2018.
%\bibitem{IEEEhanzo:noma}
%L. Dai, B. Wang, Z. Ding, Z. Wang, S. Chen, and L. Hanzo, “A survey of non-orthogonal multiple access for 5G," \emph{IEEE Commun. Surveys Tuts.}, vol. 20, no. 3, pp. 2294–2323, 3rd Quart., 2018.
%\bibitem{IEEEcover:bcast}
%H. Viswanathan, S. Venkatesan and H. Huang, "Downlink capacity evaluation of cellular networks with known-interference cancellation," \emph{IEEE Jour. on Sel. Areas in Commun.}, vol. 21, no. 5, pp. 802-811, June 2003.
%\bibitem{IEEEperotti:rateadptnoma}
%A. G. Perotti and B. M. Popović, "Non-orthogonal multiple access for degraded broadcast channels: RA-CEMA," \emph{2015 IEEE Wireless Commun. and Networ. Conference (WCNC)}, New Orleans, LA, USA, 2015, pp. 735-740.
%\bibitem{IEEElaw:precodser}
%K. L. Law and C. Masouros, "Symbol Error Rate Minimization Precoding for Interference Exploitation," \emph{IEEE Trans. on Commun.}, vol. 66, no. 11, pp. 5718-5731, Nov. 2018.
%\bibitem{IEEEzhang:precodser}
%L. Zhang, L. Gui, X. Mo and X. Sang, "Symbol Error Rate Minimization Based Constructive Interference Precoding for Multi-User Systems," \emph{IEEE Access}, vol. 9, pp. 42543-42555, 2021.
%\bibitem{IEEEagarwal:outage}
%A. Agarwal, R. Chaurasiya, S. Rai and A. K. Jagannatham, "Outage Probability Analysis for NOMA Downlink and Uplink Communication Systems With Generalized Fading Channels," \emph{IEEE Access}, vol. 8, pp. 220461-220481, 2020.
%\bibitem{IEEEwng:constnoma}
%Y. Wang, T. Zhou, T. Xu and H. Hu, "A Rotated-Constellation Based Method for BER Analysis in Uplink NOMA Systems," \emph{IEEE Trans. on Veh. Tech.}, vol. 72, no. 2, pp. 2632-2637, Feb. 2023.
%\bibitem{IEEElee:mqamser}
%C-J Kim, Y-S Kim, G-Y Jeong, J-K Mun, H-J Lee, “SER analysis of QAM with space diversity in Rayleigh fading channels,” \emph{ETRI Journ.} vol. 17, no. 4, pp. 25–35, 1996
%\bibitem{IETsha:ccdma}
%I. L. Shakya, F. H. Ali, and E. Stipidis, “High user capacity collaborative code-division multiple access”.  \emph{IET Commun.}, vol. 5, no. 3, 307–319, 2011.
%\bibitem{IEEEsha:JD-NOMA}
%I. Shakya and F. Ali, ``Collaborative Signal Multiple Access: A Robust Alternative to NOMA Using SIC'', TechRxiv. Preprint. https://doi.org/10.36227/techrxiv.14330453.v1, 2021.
%\bibitem{Isha:colla}
%I. L. Shakya,  \emph{High Capacity CDMA and Collaborative Techniques}, PhD Thesis, University of Sussex, 2008
%\bibitem{IEEEwalid:groupnoma}
%W. A. Al-Hussaibi and F. H. Ali, "Efficient User Clustering, Receive Antenna Selection, and Power Allocation Algorithms for Massive MIMO-NOMA Systems," in \emph{IEEE Access}, vol. 7, pp. 31865-31882, 2019.
%\bibitem{IEEElee:JD}
%J. Lee, D. Toumpakaris and W. Yu, “Interference Mitigation via Joint Detection," \emph{IEEE J. Sel. Areas Commun.}, vol. 29, no. 6, pp. 1172-1184, June 2011.
%\bibitem{MGProakis:DGCom}
%J. G. Proakis, and S. Masoud,  \emph{Digital Communications}, Boston: McGraw-Hill, 2008.
%\bibitem{DTse:WCom}
%D. Tse and P. Viswanath, \emph{Fundamentals of Wireless Communication}. Cambridge university press, 2005.
\bibitem{GoldS:WCom}
A. Goldsmith, \emph{Wireless Communications}. Cambridge, U.K.: Cambridge
Univ. Press, 2005.

\end{thebibliography}
\end{document}